\documentclass[twocolumn,showpacs,amsmath,amssymb]{revtex4}
\usepackage{graphicx}
\usepackage{slashed}
\usepackage{pifont}
\usepackage{color}

\newcommand{\E}[1]{\times 10^{#1}}
\newcommand{\beqn}{\begin{equation}}
\newcommand{\eeqn}{\end{equation}}

\newcommand{\veps}{\varepsilon}
\newcommand{\tr}{\mathrm{tr}\,}

\newcommand{\defn}{:=}

\def\lambbar{{\mathchar'26\mkern-9mu\lambda}}

\newcommand{\Leff}{\ensuremath{V_{\mathrm{eff}}}}
\newcommand{\feff}{\ensuremath{f_{\mathrm{eff}}}}

\newcommand{\Tmm}{\ensuremath{\mathcal{T}}}
\newcommand{\Tmn}{\ensuremath{T_{\mu\nu}}}
\newcommand{\Tumn}{\ensuremath{T^{\mu\nu}}}

\newcommand{\gumn}{\ensuremath{g^{\mu\nu}}}
\newcommand{\Fmn}{\ensuremath{F^{\mu\nu}}}
\newcommand{\Kmn}{\ensuremath{K^{\mu\nu}}}

\newcommand{\cond}{\ensuremath{\langle \bar\psi  \psi \rangle}}

\begin{document}
\title{QED Energy-Momentum Trace as a Force in Astrophysics}
\author{Lance Labun and Johann Rafelski}
\affiliation{Department of Physics, University of Arizona, Tucson, Arizona, 85721 USA, and\\
  Department f\"ur Physik der Ludwig-Maximillians-Universit\"at M\"unchen und\\
  Maier-Leibniz-Laboratory, Am Coulombwall 1, 85748 Garching, Germany}

\date{15 February, 2010}

\begin{abstract} 
We study the properties of the trace $\Tmm$ of the QED energy-momentum tensor in the presence of quasi-constant external electromagnetic fields.  We exhibit the origin of $\Tmm$ in the quantum nonlinearity of the electromagnetic theory.  We obtain the quantum vacuum fluctuation-induced interaction of a particle with the field of a strongly magnetized compact stellar object.
\end{abstract}


\maketitle
\subsection*{Introduction}
Quantum electrodynamics (QED) in (quasi-)constant, homogeneous external 
electromagnetic (EM) fields provides an opportunity to study the properties 
of the vacuum state structure under the conditions of extreme external fields.  
In the presence of an electromagnetic field that varies negligibly on the 
space-time scale of the electron-positron fluctuations in the vacuum 
$\lambbar_c=\hbar /m_ec$ leads to an effective nonlinear electromagnetic 
theory via the Euler-Heisenberg (EH) effective action%
~\cite{EKH,Schwinger:1951nm,Brezin:1970xf,BialynickaBirula:1970vy,Adler:1971wn,%
Muller:1977mm,Dittrich:2000zu, Cho:2000ei,Jentschura:2001qr,Dunne:2004nc}. 
The physical observables and effective action induced by quasi-constant external 
electromagnetic fields are well-defined, because QED is an infrared-stable theory 
in which the electron mass $m_e$ is the key scale parameter.

The requirement of Lorentz symmetry admits only one essentially new 
contribution in the final expression for the energy-momentum tensor 
$\Tumn$, a vacuum energy term proportional to $\gumn$.  This term 
is similar to Einstein's form of dark energy, the cosmological 
constant $\Lambda$.  The related repulsive anti-gravity like effect 
of the energy of vacuum fluctuations may become accessible to 
laboratory experiments: pulsed laser technology is advancing 
rapidly towards the `critical' field strength 
$$ E_c = B_c = m^2/e=1.3\E{18}\mathrm{\,V/m}\,=\,4.4\E{9}\,\mathrm{T},$$ 
making strong-field QED effects possible in the laboratory~\cite{Mourou:2006zz}.  
Magnetic fields of strength $B\gtrapprox B_c$ are also encountered 
in the study of supernovae and post-main sequence stellar 
objects~\cite{Ardeljan05,Woosley06,Peng07,Janka:2006fh}, 
and we show here how the vacuum energy leads to a novel often repulsive
force between the stellar magnetic field and a charged particle, 
influencing matter accretion and the supernova bounce.

Much of what we present here is a general property of any nonlinear 
theory of electrodynamics~\cite{Labun:2008qq}, applicable also to 
Born-Infeld electromagnetism, for example.  We achieve considerable 
simplification and insight exploiting a common feature of any 
nonlinear electromagnetism, namely the presence of a dimensioned 
field scale $E_c=m^2/e$ which we express using a mass scale $m$.  
We write the (nonlinear) effective electromagnetic action
\begin{equation}\label{Veff}
\overline\Leff \equiv -\mathcal{S}+
m^4\,\overline\feff\left(\frac{\mathcal{S}}{m^4},\frac{\mathcal{P}}{m^4}\right),
\end{equation}
\begin{equation}
\mathcal{S}\defn \frac{1}{4}g_{\kappa\alpha}g_{\lambda\beta}F^{\alpha\beta}F^{\kappa\lambda}
     =\frac{1}{2}(B^2-E^2)
\end{equation}
\begin{equation}
\mathcal{P}\defn \frac{1}{4}g_{\kappa\alpha}g_{\lambda\beta}\widetilde F^{{\alpha\beta}}F^{\kappa\lambda}
         =E\cdot B
\end{equation}
as a function of the (Lorentz) scalar  $\mathcal{S}$ and pseudo scalar $\mathcal{P}$.
In Eq.\,\eqref{Veff}  $\overline\feff$ contains solely 
contributions nonlinear in $\mathcal{S}$, excluding a possible  linear term, e.g. 
$\mathcal{S}\ln m/\mu $ where $\mu$ is another scale. When such  term 
is included we omit the bar on $\feff$.  

\subsection*{Energy-Momentum Trace}
We study the energy-momentum tensor for an effective 
nonlinear action ($g= \mathrm{det}\,g_{\mu\nu} $)
\begin{align}
\Tumn\:&=
\frac{-2}{\sqrt{-g}} \frac{\delta}{\delta g_{\mu\nu}}\int d^4x\sqrt{-g} \:\Leff
  \\
&= \veps\Tumn_{\rm Max}
 -\gumn\left(\Leff-\mathcal{S}\frac{\partial \Leff}{\partial \mathcal{S}}
               -\mathcal{P}\frac{\partial \Leff}{\partial \mathcal{P}} \right).
  \label{SchTmn}
\end{align}
The dimensionless dielectric function $\veps=-{\partial\Leff}/{\partial\mathcal{S}}\to 1$ 
in the the classical Maxwell $\Tumn_{\rm Max}$ limit $\Leff\to -\mathcal{S}$.  
The form of Eq.\,(\ref{SchTmn}) agrees with 
Eq.\,(A3)~in~\cite{Grundberg:1994kk} and Eq.\,(4.17)~in~\cite{Shore:1995fz}.
 
The first term in Eq.\,(\ref{SchTmn}) is traceless. 
The second term in Eq.\,(\ref{SchTmn})  provides  as noted 
the only possible covariant extension and is identically 
the trace of energy momentum tensor.  Using Eq.\,\eqref{Veff}
\beqn \label{dVdm}
T_\mu^\mu\!\equiv\! \Tmm \!
   =\!-4\left(\!\Leff-\mathcal{S}\frac{\partial \Leff}{\partial \mathcal{S}}
           -\mathcal{P}\frac{\partial \Leff}{\partial \mathcal{P}}\!\right)\!=\!
           - m\frac{d \overline\Leff}{dm}.
\eeqn
Our separation of the trace $\Tmm$  from the off-diagonal Maxwell-like part 
of the energy-momentum tensor isolates the field induced, gravitating
energy-momentum of the vacuum.

Before exploring the physical consequences, we must pause and clarify the precise meaning of \Tmm. 
Terms linear in the invariant $\mathcal{S}$ do not contribute 
to the right side of Eq.\,\eqref{dVdm} since they cancel explicitly 
in the middle parentheses.  Only nonlinear (in $\mathcal{S}, \mathcal{P}$) 
EM-theories can have an energy-momentum trace and the bar above \Leff~reminds 
us of this.  This is another way to say that QED with massive electrons 
is \emph{ab initio} not conformally  symmetric. Massive QED does not share the more 
challenging issues, such as conformal 
symmetry breaking, surrounding parallel efforts in quantum chromodynamics 
(QCD)~\cite{Schutzhold:2002pr}.  

A better understanding of these remarks is achieved in QED by connecting 
$\Tmm$ to the Dirac (electron-positron) condensate induced in the 
vacuum~\cite{Dittrich:1996dz}, which is directly related to the effective action
\beqn \label{dLdm-prop} 
-m\cond=im\,\tr(S_F -S_F^0)=m\frac{d\Leff}{d m}. 
\eeqn
The middle expression of Eq.\,(\ref{dLdm-prop}) exhibits the condensate 
as the difference between normal ordering of operators in the no-field 
(also called perturbative) vacuum and the with-field vacuum.
The argument of the derivative in Eq.\,(\ref{dLdm-prop}) and 
in Eq.\,(\ref{dVdm}) is not exactly the same: 
The difference in format between Eq.\,(\ref{dVdm}) and  Eq.\,(\ref{dLdm-prop})
is due to the leading linear term in $ \mathcal{S}$
\beqn\label{linearCor2} 
m\frac{d\Leff}{dm}=m\frac{d\overline\Leff}{dm}+\frac{2\alpha}{3\pi}\mathcal{S}.
\eeqn

Combining Eqs.\,\eqref{dVdm},\,\eqref{dLdm-prop}, and \eqref{linearCor2},
we obtain the well-known relation~\cite{Adler:1976zt}
\begin{align}
\label{QED-Tmm}
\Tmm \! &= \frac{2\alpha}{3\pi}\mathcal{S} + m\cond\\
 &= \frac{2\alpha^2}{45m_e^4}(7\mathcal{P}^2\!+4\mathcal{S}^2) 
     +  \mathcal{O}(\alpha^3)\label{EHAno1pert}
\end{align}
Eq.\,\eqref{QED-Tmm} displays two contributions to the trace 
of the energy-momentum tensor: gauge field and matter field fluctuations. 
There is an important exact cancellation between these two terms. 
Were the first term in Eq.\,\eqref{QED-Tmm} (erroneously) omitted, 
the trace $\Tmm$ would be that much greater, and  for any applied 
magnetic field the ${00}$-energy-density-component of the 
energy-momentum tensor would  be negative.  The QED vacuum 
would be unstable, and the naive perturbative QED vacuum
could  reduce its energy by  spontaneously generating  a state with
magnetic field.  For all practical purposes, 
the form of Eq.\,\eqref{QED-Tmm} is confirmed by the observed stability of the
QED vacuum and work claiming otherwise will need to address that important issue. 

In fact, the relative sign in Eq.\,\eqref{QED-Tmm} agrees with
Eq.\,(5) of Ref.~\cite{Schutzhold:2002pr} and  Eqs.(35) in \cite{Gorbar:1999xi}
with the recognition that the (fermion) Gell-Man-Low $\beta$-function in QED is
positive definite.
Our Eq.\,\eqref{QED-Tmm} (and the more explicit form of \Tmm, 
Eq.\,\eqref{EHAno1} below), agrees with~\cite{Dittrich:1996dz}, 
which result is a bit surprising since it follows
from the clearly contradictory  supposition that $\Tmm = m\partial\Leff/\partial m$. 
Moreover, there are quite a few other instances in literature where the first term 
in Eq.\,\eqref{QED-Tmm} is omitted.

\subsection*{Strong  Fields Simulacrum of Dark Energy}
Applying Eq.\,\eqref{QED-Tmm} with the Euler-Heisenberg-Schwinger effective 
action (using Schwinger's notation and units in which 
$\alpha = e^2/4\pi$~\cite{Schwinger:1951nm}) gives an explicit formula for the trace,   
\beqn
\Tmm    = \label{EHAno1} 
\frac{2\alpha}{3\pi} \mathcal{S} - 
 \frac{m^2}{4\pi^2}\!\int_{0}^{\infty}\!\! \!\!ds\,e^{-m^2s}
  \left(\!e^2ab\frac{\coth(ebs)}{\tan(eas)}-\frac{1}{s^2}\!\right) 
 \eeqn
wherein the invariant magnetic- and electric-like field strengths are
$$b^2= \sqrt{\mathcal{S}^2+\mathcal{P}^2}+\mathcal{S} \to B^2 , \ 
     a^2=\sqrt{\mathcal{S}^2+\mathcal{P}^2}-\mathcal{S} \to E^2, $$
reducing as indicated to the classical magnetic and electric fields 
when one invariant vanishes.

\begin{figure}[!tb]
\includegraphics[width=.48\textwidth]{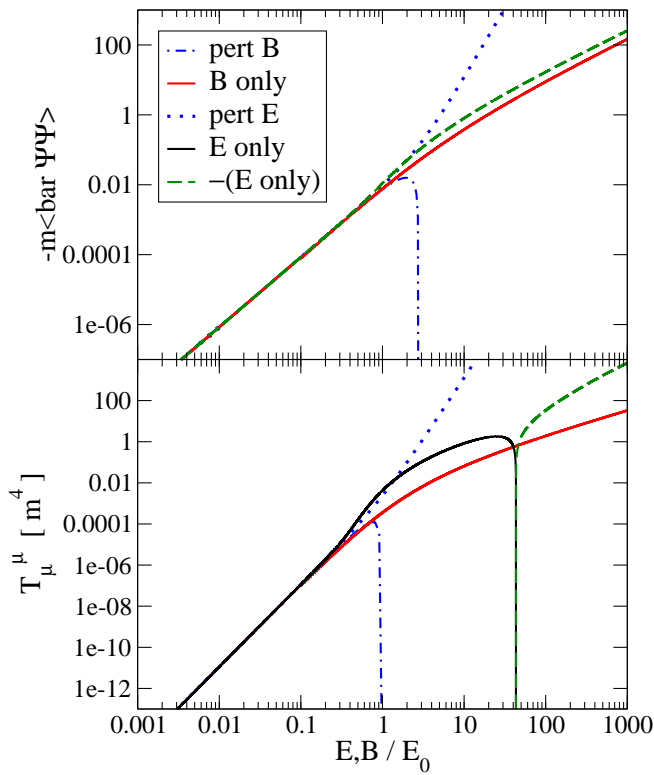}
\vspace*{-0.4cm}
\caption{The condensate $-m\cond$ (top) and the trace of energy-momentum tensor
$ T_{\mu}^{\:\:\mu}$ (bottom) in units of $m^4$,
as a function of magnetic $B$ (red) and electric $E$ (blue, dashed and solid)
field strengths. The negative of the electric field result is plotted where
appropriate.  The dotted (for $E$) and dashed-dotted (for $B$) 
lines show  the weak-field expansions up to $E^6,B^6$.
\label{spin-cond}   }
\end{figure}

In numerical evaluation of the  energy-momentum tensor for arbitrarily 
strong fields we employ the method developed in~\cite{Muller:1977mm}. 
Consider first the stable field configuration $B\ne 0,E=0$.  
The subtracted meromorphic (i.e. residue) expansions of the function
\beqn\label{Mer1}
 x \coth x - 1=\!\! \sum_{k=1}^{\infty}\frac{2x^2}{x^2+k^2\pi^2}
  =\frac{x^2}{3} -\!\sum_{k=1}^{\infty}\frac{1}{(k\pi)^2}\frac{2x^4}{x^2+k^2\pi^2}. 
\eeqn
display the stabilizing change in sign following the second subtraction.  
The sums and integrals are absolutely convergent, so we may resum the resulting series, 
obtaining
\beqn\label{Leff-Trep}
\Leff(B) = 
\frac{m^4}{16\pi^2\beta'}\int_0^{\infty}\!\!\!dz\,\ln(z^2+1)\ln(1-e^{-\beta'z}),
\eeqn
\beqn \label{Con-step2} 
-m\cond_B = 
-\frac{m^4}{2\pi^2\beta'}\int_0^{\infty}\frac{\ln(1-e^{-\beta' z})}{1+z^2} dz>0,
\vspace*{-0.35cm}
\eeqn
\beqn \label{Tmm-B-step2}
\Tmm_B =
  -\frac{m^4}{2\pi^2\beta'}
  	\int_{0}^{\infty} \frac{z^2\ln(1-e^{-\beta' z})}{1+z^2}dz>0,
\eeqn
in which $\beta' = \pi m^2/eB=\pi/(B/E_c)$.  
Eq.\,\eqref{Leff-Trep} presents the (renormalized) 
effective action also seen in Ref.~\cite{Muller:1977mm}.  
Numerical evaluations of the condensate  Eq.\,\eqref{Con-step2}
and trace Eq.\,\eqref{Tmm-B-step2} in the presence of magnetic 
and electrical field (real part only) are shown in figure~\ref{spin-cond}.
We discuss elsewhere~\cite{Labun:2008qq} the case that both $E,B$ are non zero,
how real and imaginary parts contribute together for electric fields,  
and the case of spin-0 matter fields. 
For fields way above critical the results presented in  figure~\ref{spin-cond}
are not accessible in practice by direct integration of
the  proper time representation   Eq.\,\eqref{EHAno1}

For the electric field case, the corresponding meromorphic 
expansions of $x\:{\rm cot}\:x$ show poles on the real $s$-axis. 
We assign to the mass a small imaginary component $m^2\to m^2+i\epsilon$, 
replacing in Eq.\,(\ref{Con-step2}) and Eq.\,(\ref{Tmm-B-step2}) 
the denominator $z^2+1 \to z^2-1+i\epsilon$.  
Thus, in the presence of an electric field, 
there is also nonperturbative imaginary contribution to \Tmm.

From now on in this work we address strong magnetic fields.
$\Tmm$ in the presence of a magnetic field is positive for any 
given field $B$, in contrast to the negative of the condensate $m\cond$. 
The manifest signs of the two expressions Eqs.\,\eqref{Con-step2} 
and \eqref{Tmm-B-step2}, which determine the physics 
outcome of this investigation justify the time and effort spent showing
 how \Tmm~does not include the term linear in $\mathcal{S}$, while $m\cond$  
does. Clarification of this exclusion is necessary since as noted 
$\Tmm$ and $m\cond$ are often conflated in literature.

The trace $\Tmm$ gravitates, just as the Casimir energy does~\cite{Fulling}.
Because in the Euler-Heisenberg-Schwinger calculation the `constant' external 
field is global in extent, this energy-momentum is manifested in the form of 
a cosmological constant.  In contrast to matter, for which the particle 
pressure acts outwards, the pressure part of energy-momentum tensor described 
by $-\Tmm/4$ acts inwards.  This is a general feature of any  
`false' vacuum state: the outside true vacuum  the squeezes the false, 
higher energy density vacuum out of existence.

The sign reversal of $\Tmm$ pressure (compared to pressure of regular matter) 
overwhelms the gravity of the positive energy density, providing the 
anti-gravity effect associated with dark energy.  The similarity of $\Tmm$ 
to the cosmological constant was noted before by Sch\"utzhold~\cite{Schutzhold:2002pr} 
and can be made explicit in the Einstein equation by separating 
the trace, $\widetilde\Tmn=\Tmn-g_{\mu\nu}\Tmm/4$
\beqn \label{tr+EEqs1}
\frac{1}{8\pi G}\left(R_{\mu\nu}-\frac{1}{2}g_{\mu\nu} R\right) 
   = -\widetilde\Tmn -g_{\mu\nu}\left(\frac{\Tmm}{4} +\frac{\lambda}{2}\right).
\eeqn

With a sign like that of dark energy 
$\Lambda/4\pi G\equiv{\lambda}\simeq (2.3$\,meV$)^4=4.1\E{-34}m_e^4$, 
\Tmm~is the dominant contribution in 
a domain of space with strong fields and is naively expected to generate 
a pressure that sweeps out matter, in analogy with the cosmological 
acting constant at large scales and pushing the universe apart.  
For comparison, we note that the magnetic field-induced vacuum 
energy rivals cosmological dark energy $\lambda$ at $B_d=108$\,T,
 the scale of the largest static laboratory fields.

Energy-momentum sources in Eq.\,\eqref{tr+EEqs1} 
involving local dark energy-like contributions have only recently been
studied~\cite{Mazur04,Chapline:2005ph,Lobo:2005uf,Chan:2008rk}.
To see that the effects of local $\Tmm$ are anti-gravitational just 
like cosmological $\lambda$, we inspect
the Oppenheimer-Volkov equation 
\beqn\label{OV}
\frac{dp}{dr} = 
-\frac{G}{c^2}(T^0_0 + T^i_i)\frac{M+4\pi r^3T^i_i}{r(r-2GM)}
\eeqn
where $M(r) = \int^r T^0_04\pi r'^2dr'$ and $T^i_i=p-\lambda$ is assumed isotropic.  
$\lambda$ does not contribute to the first term which is always 
non-negative: $T^0_0 + T^i_i=(\rho+\lambda)+(p-\lambda)$, 
but $\lambda$ can make the second term 
$(M+4\pi r^3T^i_i)$ change sign.  Contributions to $T^{\mu}_{\nu}$ 
proportional to $g^{\mu}_{\nu}$ (like $\lambda$) thus weaken the pressure 
gradient and support heavier stars than otherwise expected.

We have checked using Eq.\,\eqref{OV} that direct gravitational 
modifications to the mass-radius relation for compact stellar 
objects remain negligible, as might be expected.
At $60B_c$, $\Tmm=0.8 m_e^4=1.14\E{25}$erg/cm$^3$, 
8 orders of magnitude smaller than the pressures expected 
in the high density nuclear matter in the core 
of a post-main sequence star~\cite{Janka:2006fh}.
However, this energy density is 2-4 orders of magnitude larger than 
the gravitational potential energy density of infalling stellar plasma.  
Thus while the insignificance of the QED $\Tmm$ in gravity is a 
consequence of its objectively small energy density,  the anti-gravitational 
effect of $\Tmm$ suggests a closer study of the force experienced by 
individual particles in nonlinear electromagnetism is necessary.

\subsection*{Particles in Overcritical Quasi-Constant Fields}
Forces present in the dynamical case can be much greater than 
those observed when the interacting bodies are studied in the 
hydrostatic equilibrium of Eq.\,\eqref{OV}.  The electromagnetic 
force determining individual particle dynamics does not make 
relevant contributions, and the 
{\em vacuum fluctuation-induced force} contributes even 
less to  Eq.\,\eqref{OV}. Though very much smaller than Maxwell's 
force this force can be stronger than gravity and at times more 
relevant than the linear order force of Maxwellian electromagnetism. 
We will describe its features relevant to the charged  particle 
dynamics within collapsing stellar objects.

Consider that the total (`t') electromagnetic energy-momentum 
tensor $\Tumn_{\rm t}$ due to both an external field (`e') and 
a probe charge (`p') includes also an interaction 
energy-momentum $\Tumn_{\rm int}$, 
\beqn\label{intTmn-defn}
\Tumn_{\rm t} = \Tumn_{\rm e}+\Tumn_{\rm p}+\Tumn_{\rm int}
\eeqn
with tensors $\Tumn_{\rm e},\Tumn_{\rm p}$ defined by the forms 
they take in isolation from each other. When the external field 
is much larger than the field of the probe particle, the 
electromagnetic energy-momentum tensor is expanded in the 
displacement tensor $\Kmn$ 
\beqn\label{constitutive}
\Kmn =-\frac{\partial\Leff}{\partial F_{\mu\nu}}
     =\Fmn-\frac{\partial\overline{f_{\rm eff}}}{\partial F_{\mu\nu}}.
\eeqn
around the dominant contribution of the external field,
\beqn\label{Tmn-exp}
\Tumn_{\rm t}=
\Tumn_{\rm e}+
\left.\frac{\partial\Tumn}{\partial K^{\alpha\beta}}\right|_{\rm e}\!
  \frac{K_{\rm p}^{\alpha\beta}}{2}+...
\eeqn
with the subscript ${\bf e}$ reminding that derivatives are 
to be evaluated at the external field.  

The energy-momentum tensor $\Tumn$ is expressed in 
Eq.\,\eqref{SchTmn} in terms of the field tensor $\Fmn$, 
but only the displacement fields of the probe particle are 
known explicitly by solving Maxwell's equations with source  
\beqn\label{sourcedMax}
\partial_{\mu}\Kmn_{\rm p}=j^{\nu}_{\rm p}\,.
\eeqn
By inverting Eq.\,\eqref{constitutive} we obtain
\beqn\label{dFdK}\begin{split}
\frac{\partial F^{\alpha\beta}}{\partial \Kmn}=&
(\delta^{\alpha}_{\mu}\delta^{\beta}_{\nu}
-\delta^{\alpha}_{\nu}\delta^{\beta}_{\mu})
+\frac{\partial^2 \overline{f_{\rm eff}}}{\partial \Fmn\partial F_{\alpha\beta}} \\ 
&\hspace*{1cm} 
+\frac{\partial^2 \overline{f_{\rm eff}}}{\partial \Fmn\partial F_{\gamma\delta}}
\frac{\partial^2 \overline{f_{\rm eff}}}{\partial F^{\gamma\delta}\partial F_{\alpha\beta}}
+\ldots\,
\end{split} \eeqn
This  rank 4 tensor transforms the functional dependence from the field tensor 
to the displacement tensor. We checked the validity of truncation by numerical 
evaluation of the derivatives of the action, which shows that (normalized) higher 
derivatives are suppressed even when the field is supercritical. A separable 
contribution of the Maxwell self-energy $\Tumn_{\rm p}\approxeq\Tumn_{\rm Max,p}$ 
of the probe particle is indeed found at next order $\partial^2/\partial K^2$ and 
would be subtracted.  However, we find the effects of the terms from the second 
derivative are many orders of magnitude smaller than those from the first 
derivative and do not discuss them further here.

Using Eq.\,\eqref{dFdK} in Eq.\,\eqref{Tmn-exp} gives 
\beqn\label{dTdH}
\Tumn_{\rm int} = \Tumn_{\rm ep}+(\veps-1)^2 \Tumn_{\rm ep} 
                   +g^{\mu\nu}\frac{2-\veps}{4}\frac{\partial\Tmm}
{\partial\mathcal{S}}F^{\rm e}_{\alpha\beta}K_{\rm p}^{\alpha\beta}
\eeqn
where
\beqn\label{Tmnep}
\Tumn_{\rm ep} = -(F_{\rm e}^{\mu\kappa}K_{\rm p}^{\nu\lambda}+F_{\rm e}^{\nu\lambda}
K_{\rm p}^{\mu\kappa})g_{\kappa\lambda}+g^{\mu\nu}
\frac{1}{2}F^{\rm e}_{\alpha\beta}K_{\rm p}^{\alpha\beta}
\eeqn 
is the  Maxwellian interaction with 
$F^{\rm e}_{\alpha\beta}K_{\rm e}^{\alpha\beta}
=2(\vec B_{\rm e}\!\cdot\! \vec H_{\rm p}
-\vec E_{\rm e}\!\cdot\!\vec D_{\rm p})$. 
The latter two terms of Eq.\,\eqref{dTdH} remain after 
cancellation among the order $\alpha$ terms, and despite 
being order $\alpha^3$ the last term $(1-\veps)(\partial\Tmm/\partial\mathcal{S})
=(\partial\overline{f_{\rm eff}}/\partial\mathcal{S})(\partial\Tmm/\partial\mathcal{S})$ 
is kept for now.

We view the net force (density) acting on the charged probe particle 
entering the domain of the external field in the usual way, requiring 
that inertial resistance balance any breach of the conservation of 
the field energy-momentum, 
\beqn\label{cov-f-defn}
f^{\mu}\equiv-\partial_{\nu}\Tumn_{\rm int}
    =j_{\nu}F_{\rm e}^{\mu\nu}+\delta f^{\mu}
\eeqn
using that $-\partial_{\nu}\Tumn_{\rm ep}=j_{\nu}^{\rm p}F^{\mu\nu}_{\rm e}$ 
is the force obtained within Maxwell's linear electromagnetism.  Here
\beqn\label{delta-4force}\begin{split}
\delta f^{\mu} \!&\approxeq\! (\veps_{\rm e}\!-1)^2j_{\nu}^{\rm p}F^{\mu\nu}_{\rm e} \!
-\Tumn_{\rm ep}\partial_{\nu}(\veps_{\rm e}\!-1)^2\! \\[0.2cm] 
&\hspace*{0.5cm}
-\partial^{\mu}\!\left.\frac{2\!-\!\veps}{4}
\frac{\partial\Tmm}{\partial\mathcal{S}}\right|_{\rm e}\!\! 
F^{\rm e}_{\alpha\beta}K_{\rm p}^{\alpha\beta}
\end{split}\eeqn
with equality only approximate on account of the finite order expansions 
in $\Tumn$ Eq.\,\eqref{Tmn-exp} and $\alpha$ Eq.\,\eqref{dFdK}.  

The conventional contributions of 
$\Tumn_{\rm ep}$ and $F^{\rm e}_{\alpha\beta}K_{\rm p}^{\alpha\beta}$, 
i.e. the first term  of Eq.\,\eqref{dTdH}, to the force on the 
particle are obtained by integrating over a covariant hypersurface, 
and all frames being equivalent this integration is done most 
conveniently in the rest frame of the particle, 
\beqn\label{EdotD-int}
\int \!d^3x\: E_{\rm e}^iD_{\rm p}^j =
 -\!\int \!d^3x\: \Phi_{\rm e}\,\nabla^i\nabla^j\Phi_{\rm p} 
= \delta^{ij}\!\int\! d^3x\:\Phi_{\rm e}\rho_{\rm p}
\eeqn
using that $\vec D_{\rm p}=(Ze/r^2)\hat r$ (for a spherical charge) 
has only one component when choosing spherical coordinates.  

A physical charged particle has also a magnetic moment $\vec \mu$, 
and thus a corresponding dipole magnetic field. As is well known 
this part of the force cannot usually compete with the effect of 
electrical particle charge. However, in the present context we 
reach beyond the usual Lorentz force to the effect of vacuum 
fluctuations and it is necessary to check if it is still justified 
to neglect the magnetic dipole in the strong magnetic field environment 
of a collapsing star. From the magnetic dipole interaction energy 
$\vec\mu_{\rm p}\!\cdot\!\vec B_{\rm e}$ a force due to the gradient 
of the external magnetic field arises.  $B_{\rm e}$ changes on 
macroscopic scale though, and the parameter characterizing smallness 
of the effect is $1/mL\simeq 10^{-16}$ when $B_{\rm e}$ varies on the 
scale $L\simeq 4$\,km.  For comparison, the smallness parameter of the 
vacuum fluctuations arising from Euler-Heisenberg action is 
$(B_{\rm e}/B_c)^2\alpha/45\pi$.  Seeing that vacuum fluctuation effects 
should dominate the magnetic dipole interaction for a stellar 
magnetic field $B_{\rm e}>10^{-5}B_c$, we explore this domain further.

Turning now to the latter two terms of Eq.\,\eqref{dTdH} that represent
the additional vacuum fluctuation-induced force, we observe that 
the gradient in the corresponding last term of Eq.\,\eqref{delta-4force} 
generates two contributions: the first as the gradient of 
Eq.\,\eqref{EdotD-int} and the second as the net change of the 
slowly-varying coefficient $\partial\Tmm/\partial\mathcal{S}$ 
over the domain of particle's field.
The integrals over the particle's field Eq.\,\eqref{EdotD-int} computed 
in the particle's rest frame,  the spatial components of the force 
Eq.\,\eqref{delta-4force} on a point charge 
$\rho_{\rm p}=Ze\delta(\vec x)$ in its rest frame are
\beqn
\label{delta-3f}
\frac{1}{Ze}\delta \vec f \approxeq
\left.\left((\veps-1)^2+\frac{\veps-2}{2}
\frac{\partial\Tmm}{\partial\mathcal{S}}\right)\right|_{\rm e}\!\vec E_{\rm e}
-C_{\rm e}\Phi_{\rm e}\vec \nabla \mathcal{S}_{\rm e},
\eeqn
where 
\beqn
\label{A-coeff} 
C_{\rm e}B_c^2 
=-\!\left(\!2(\veps-1)\frac{\partial\veps}{\partial\mathcal{S}} 
+\frac{\veps-2}{2}\frac{\partial^2\Tmm}{\partial\mathcal{S}^2}
+\frac{1}{2}\frac{\partial\veps}{\partial\mathcal{S}}
\frac{\partial\Tmm}{\partial\mathcal{S}}\!\right)
\eeqn
depends only on the scalar invariant of the external field $\mathcal{S}_{\rm e}$. 
On the right side of Eq.\,\eqref{delta-3f} the gradient 
applied to the invariant $\mathcal{S}$ preserves the correct 
Lorentz transformation property: although the potential 
$\Phi_{\rm e}$ appears, Eq.\,\eqref{delta-3f} is the gauge 
invariant correction to the linear force 
$\vec f = e(\vec E_{\rm e}+\vec v \times \vec B_{\rm e})$ 
computed in the particle's rest frame.

The component of Eq.\,\eqref{delta-3f} proportional to 
$C_{\rm e}$ is qualitatively different
because, being proportional the gradient $\vec\nabla\mathcal{S}$, 
it allows the transfer of energy from the magnetic field to 
in-falling particles. This property is in contrast to the 
first term in Eq.\,\eqref{delta-3f}  which produces a 
tiny change in the effective linear force (per mille at $B_{\rm e}=B_c$). 

The weak-field expansion of $C_{\rm e}$ Eq.\,\eqref{A-coeff}
\beqn
C_{\rm e}B_c^2\approxeq \frac{8\alpha}{45\pi}\!
\left(1-\left(\!\frac{B_{\rm e}}{B_c}\!\right)^{\!2}\!\left(\frac{6}{7}
              +\frac{2\alpha}{45\pi}+\ldots\right)+\ldots\right) \label{A-coeff-exp}
\eeqn
obtained from the Euler-Heisenberg effective action, 
is usable up to $B_{\rm e}\approxeq .1B_c$.  The expansion 
shows that the predominant contribution to the gradient 
force is $\Tmm$, as the leading constant 
in Eq.\,\eqref{A-coeff-exp} is traced to $\partial^2\Tmm/\partial\mathcal{S}^2$.  
Non-perturbative computation requiring employment of the nonperturbative 
numerical methods presented of the coefficients 
$\veps-1, \partial\Tmm/\partial\mathcal{S},$\,etc. shows that 
the force persists in the considered high magnetic field domain 
despite what the perturbative expansion suggests.

\begin{figure}[!tb]
\includegraphics[width=.48\textwidth]{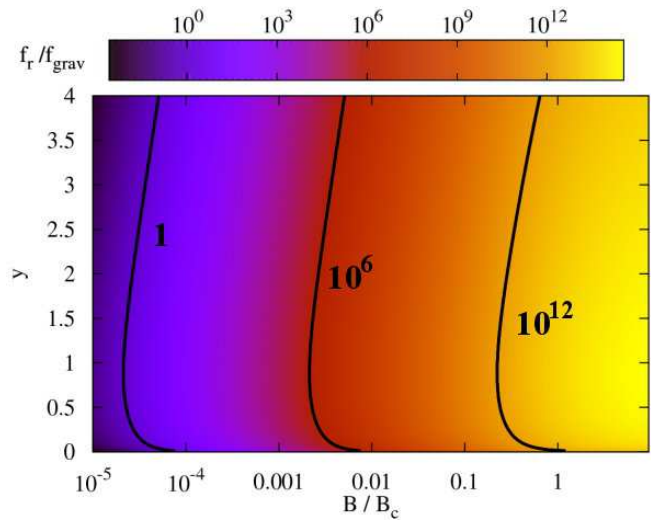}
\vspace*{-0.6cm}
\caption{The ratio $f_r/f_{\rm grav}$ Eq.\,\eqref{f-example} 
for a transverse ($\psi=\pi/2$) electron of rapidity 
$y =\cosh^{-1} \gamma$ at the surface of a $1.5 M_{\odot}$, 12 km 
radius star with magnetic field given on the horizontal axis.  
Using the expansion Eq.\,\eqref{A-coeff-exp} for $C_{\rm e}$ 
produces the depicted values up to $B\approxeq .1B_c$.  (Color online.)}
\label{fig:f_rat}
\end{figure} 

\subsection*{Particle Dynamics in a Stellar Magnetic Field}
As an application, we consider the dipole field of a strongly 
magnetized star. We are studying the force $f_{\rm r}$ in the {\em rest frame} 
of the particle and  in order to allow that the particle has a velocity  
relative to the star, we need to Lorentz-transform the field in the 
star's rest frame $B'$ oriented at an angle $\psi$ from the direction 
of the particle's motion.  As a consequence of the transformation, 
the field of the star is seen by the particle to have an electric 
component. Specifically, $v^{\mu}=\gamma(1,0,0,\beta)=(\cosh y,0,0, \sinh y)$ 
and the Lorentz transformed fields are 
$\vec B_{\rm e}=B'(\cos\psi\:\hat z -\gamma \sin\psi\:\hat x)$ and 
$\vec E_{\rm e}=-B'\gamma\beta \sin\psi\:\hat y$. Using this 
in Eq.\,\eqref{delta-3f} we obtain the  force of the star's 
field on the moving particle. Note that $\beta$, $y$  can be positive or negative.

To compare with the gravitational force $f_{\rm g}$, we must also 
Lorentz transform it to the rest frame of the moving particle.  
Although general relativistic corrections to the Newtonian 
potential are significant near the stellar objects where 
such strong magnetic fields have been inferred, the Newtonian 
force is a reasonable first estimate modified only by 
multiplicative numerical factors of order unity down to 
a few times the radius of the future neutron star.

We obtain the transformation property of the force $f_{\rm g}$ 
considering  the geodesic in the Schwarzschild metric 
(see e.g. Eq. 9.32 in~\cite{Hobson:2006se}): transforming 
to the rest frame of a relativistic particle dilates the 
proper time, multiplying kinetic and total energies by $\gamma^2$.  
The energy equation for the geodesic therefore preserves 
its form if the same  factor $\gamma^2$ is included also 
in the `potential' terms, thus giving the transformation 
$f_{\rm g} \mapsto \gamma^2 f_{\rm g}$. The ratio of the 
radial vacuum fluctuation force to the Newtonian gravitational force 
for a transversely moving $(\psi=\pi/2)$ electron is
\begin{align}\label{f-example}
\!\!\!\frac{f_{\rm r}}{f_{\rm g}} &= 
3C_{\rm e} Ze\beta\gamma B_{\rm surf} \!
\left(\!\frac{R_{\rm surf}}{r}\!\right)^{\!\!9}\!\!
\left(\!\frac{eB_{\rm surf}}{m_e^2}\!\right)^{\!\!2}\!\!
\frac{r^2}{\gamma^2 GM_{\odot}m_{\rm p}},  \end{align}
illustrated in figure \,\ref{fig:f_rat}.

The ratio $f_{\rm r}/f_{\rm g}$ Eq.\,\eqref{f-example} can 
be large as shown in figure\:\ref{fig:f_rat} due to the 
weakness of gravity, particularly for $B_{\rm e}\gg 10^{-5}B_c$. 
The stellar magnetic field contribution to the gravitating energy 
density remains relatively small, yet it affects particle dynamics 
through the coupling to moving charge, through vacuum fluctuation 
nonlinearity suppressed by $\alpha^2/m_e^4$.  Regarding sign in 
Eq.\,\eqref{f-example}, CPT symmetry of the vacuum assures that 
matter and antimatter are expelled to the same degree: the effect 
of the force is the same for a left moving electron as for a right 
moving positron, as seen by simultaneously flipping the signs 
of $Z$ and $\beta$.  
Allowing for the distribution of charges and velocities with 
respect to the orientation of the field, we recognize that in 
a random medium (plasma) half of the charged particles of each 
polarity at any given time is expelled.  

The interesting feature of the force Eq.\,\eqref{delta-3f} 
and Eq.\,\eqref{f-example} is that a magnetic field gradient 
correlates velocity and charge.  As the magnetic field curves 
trajectories of both left moving electrons and right moving 
positrons in the same direction, the noted symmetry in the 
effect of the vacuum fluctuation-induced force generates 
rotation in a net neutral plasma even while net current 
remains zero.
Therefore, in a plasma made of negatively charged electrons 
and positively charged light nucleons, the  
matter which is ejected has opposite net momentum compared 
to the matter which is attracted and thus angular momentum 
is imparted to the magnetic source due to the mass asymmetry 
between positively and negatively charged particles. \\


In summary, 
we have evaluated the trace $\Tmm$ of the QED 
energy-momentum tensor and demonstrated that its gradient
entails a significant and   often repulsive force, 
which can be large compared to  gravity, even while the 
relative energy density  of $\Tmm$ remains small.
Although the magnitude of the usual magnetic force 
$\vec v\times \vec B_{\rm e}$  is much larger than 
the  vacuum-fluctuation induced correction Eq.\,\eqref{delta-3f}, 
only the latter is relevant in consideration of energy 
transfer to the particle and escape from the gravitational potential well.
The requisite energy exchange with a magnetic field, 
seen in the gradient $\vec\nabla\mathcal{S}$, is a consequence of the 
nonlinearity of the induced vacuum fluctuations, absent in classical
Maxwellian electromagnetism.
 
The quantitative study we present in Fig.\,\ref{fig:f_rat} for the 
ratio  Eq.\,\eqref{f-example}  indicates the 
force Eq.\,\eqref{delta-4force} and Eq.\,\eqref{delta-3f}, 
should have an impact on matter accretion and stellar collapse 
dynamics in  astrophysical situations  where strong magnetic fields 
in excess of $B\gg 10^{-4} B_c =10^5$T are known to exist. 
 While treatment of the complete dynamical 
situation is beyond the scope of this work, it is easy to imagine that this force  
could help the  neutrino based transport phenomena~\cite{Burrows} to propel 
the supernova bounce.

\vspace{-0.7cm}
\acknowledgments
\vspace{-0.5cm}
This work in part supported by the DFG Cluster of Excellence
MAP (Munich Centre of Advanced Photonics), we thank Prof. D. Habs for 
hospitality. Supported  by a grant from  
the U.S. Department of Energy,  DE-FG02-04ER41318.

\end{document}